\documentclass[]{aastex}

\usepackage{emulateapj5}
\usepackage{onecolfloat}
\usepackage{graphicx} 
\usepackage{fancyheadings} 
\usepackage{ulem}
\usepackage{rotating}
\usepackage{lscape}

\begin{document}

\def\intl{\int\limits}
\def\nstat{$\approx $}
\def\perd{\;\;\; .}
\def\cmma{\;\;\; ,}
\def\ltk{\left [ \,}
\def\ltp{\left ( \,}
\def\ltb{\left \{ \,}
\def\rtk{\, \right  ] }
\def\rtp{\, \right  ) }
\def\rtb{\, \right \} }
\newcommand{\snrlim}{SNR$_{lim}$}
\newcommand{\nhi}{$N_{\rm HI}$}
\newcommand{\mnhi}{N_{\rm HI}}
\newcommand{\flls}{f_{\rm HI}^{\rm LLS}}
\newcommand{\fdla}{f_{\rm HI}^{\rm DLA}}
\newcommand{\llls}{$\ell_{\rm LLS}$}
\newcommand{\ldla}{\ell_{\rm DLA}}
\newcommand{\fnhi}{$f_{\rm HI}(N,X)$}
\newcommand{\mfnhi}{f_{\rm HI}(N,X)}
\newcommand{\Nth}{2 \sci{20} \cm{-2}}
\newcommand{\taux}{$d\tau/dX$}
\newcommand{\gz}{$g(z)$}
\newcommand{\nz}{$n(z)$}
\newcommand{\nx}{$n(X)$}
\newcommand{\omg}{$\Omega_g$}
\newcommand{\ostr}{$\Omega_*$}
\newcommand{\momg}{\Omega_g}
\newcommand{\olls}{$\Omega_g^{\rm LLS}$}
\newcommand{\odla}{$\Omega_g^{\rm DLA}$}
\newcommand{\oneut}{$\Omega_g^{\rm Neut}$}	
\newcommand{\ohi}{$\Omega_g^{\rm HI}$}
\newcommand{\olwz}{$\Omega_g^{\rm 21cm}$}
\newcommand{\ndla}{71}
\newcommand{\kms}{km~s$^{-1}$ }
\newcommand{\cm}[1]{\, {\rm cm^{#1}}}
\newcommand{\mkms}{{\rm \; km\;s^{-1}}}
\newcommand{\delv}{\Delta v}
\newcommand{\lya}{Ly$\alpha$}
\newcommand{\nv}{N\,V}
\newcommand{\ovi}{O\,VI}
\newcommand{\N}[1]{{N({\rm #1})}}
\newcommand{\sci}[1]{{\rm \; \times \; 10^{#1}}}
\newcommand{\dla}{DLA}
\newcommand{\dlas}{DLAs}


\twocolumn[%
\submitted{Accepted to ApJ: August 11, 2005}
\title{The UCSD Radio-Selected Quasar Survey for Damped \lya\ Systems}

\author{Regina A. Jorgenson\altaffilmark{1}, Arthur M. Wolfe\altaffilmark{1}, Jason X. Prochaska\altaffilmark{2}, Limin Lu\altaffilmark{3}, J. Christopher Howk\altaffilmark{4}, Jeff Cooke\altaffilmark{1,5}, Eric Gawiser\altaffilmark{6}, \& Dawn M. Gelino\altaffilmark{7}}

\begin{abstract}
As large optical quasar surveys for \dlas\ become a reality and the
study of star forming gas in the early Universe achieves statistical
robustness, it is now vital to identify and quantify the sources of
systematic error.  Because the nature of optically-selected quasar
surveys makes them vulnerable to dust obscuration, we have undertaken
a radio-selected quasar survey for \dlas\ to address this bias.  We
present the definition and results of this survey.  We then combine
our sample with the CORALS dataset to investigate the \ion{H}{1} column
density distribution function \fnhi\ of \dlas\ toward radio-selected
quasars.  We find that \fnhi\ is well fit by a power-law $\mfnhi = k_1
N^{\alpha_1}$, with $\log k_1$ = $22.90$ and $\alpha_1$ =
$-2.18^{+0.20}_{-0.26}$.  This power-law is in excellent agreement
with that of optically-selected samples at low \nhi, an important yet
expected result given that obscuration should have negligible effect
at these gas columns.
However, because of the relatively small size of the radio-selected
sample, 26 \dlas\ in $119$ quasars, \fnhi\ is not well constrained at
large \nhi\ and  the first moment of the \ion{H}{1} distribution
function, \omg, is, strictly speaking, a lower limit.  The power-law is
steep enough, however, that extrapolating it to higher column
densities implies only a modest, logarithmic increase in \omg.
The radio-selected value of $\momg = 1.15^{+0.37}_{-0.38}\times$
10$^{-3}$,  agrees well with the results of optically-selected
surveys.  While our results indicate that dust obscuration is likely
not a major issue for surveys of \dlas\, we estimate that a
radio-selected sample of $\approx$ 100 \dlas\ will be required to
obtain the precision necessary to absolutely  confirm an absence of
dust bias.


\end{abstract}   
]

\altaffiltext{1}{Department of Physics, and 
Center for Astrophysics and Space Sciences, 
University of California, San Diego, 
Gilman Dr., La Jolla; CA 92093-0424; regina@physics.ucsd.edu, awolfe@ucsd.edu}

\altaffiltext{2}{Department of Astronomy and Astrophysics, 
UCO/Lick Observatory;
University of California, 1156 High Street, Santa Cruz, CA  95064;
xavier@ucolick.org}

\altaffiltext{3}{Lucent Technologies, Naperville, IL.}

\altaffiltext{4}{Dept. of Physics, University of Notre Dame, Notre Dame, IN 46556}

\altaffiltext{5}{Center for Cosmology, University of California,
Irvine, Irvine CA 92697-4575; cooke@uci.edu}

\altaffiltext{6}{NSF Astronomy \& Astrophysics Postdoctoral Fellow,
Yale Astronomy Department and Yale Center for Astronomy \& Astrophysics, PO
Box 208101, New Haven, CT  06520-8101}

\altaffiltext{7}{Michelson Science Center, Caltech, MS 100-22, 770
South Wilson Avenue, Pasadena, CA 91125}

\pagestyle{fancyplain}
\lhead[\fancyplain{}{\thepage}]{\fancyplain{}{Jorgenson et al.}}
\rhead[\fancyplain{}{The UCSD Radio-Selected Quasar Survey for \dlas}]{\fancyplain{}{\thepage}}
\setlength{\headrulewidth=0pt}
\cfoot{}

\section{Introduction}

The introduction of large data sets from surveys such as the Sloan
Digital Sky Survey (SDSS) has made possible statistically significant
studies of Damped \lya\ systems (\dlas\ )
\citep[][hereafter PHW05]{pro05}, quasar absorption systems defined as
having an \ion{H}{1} column density $\mnhi \geq \Nth$ and which
contain most of the neutral gas in the  redshift interval z=[0, 5]
\citep{wolfe05}.  These large surveys for \dlas\ shed light on the
history of the neutral gas content of the Universe and show how it is
affected by star formation and gas replenishment.  However, one major
problem consistently affects magnitude-limited optical surveys: the
issue of intervening dust and the possibility of obscuration bias.
Because the metallicities of \dlas\ can be as high as 1/3 solar and are
always above 1/1000 solar \citep{pro03},
the presence of dust in these objects is not surprising.
Specifically, evidence from element abundance patterns suggests the
presence of depletion \citep{pettini99}, while evidence for
differential reddening suggests that dust obscuration is possible
\citep{pei95}.  Dust obscuration in optically-selected surveys could
be introducing a selection bias against \dlas\ whose high dust optical
depth would hide the background quasar.  This effect could seriously
impact the results of statistics and derived values like \omg , the
cosmological density of neutral gas (see \cite{pei95}), particularly
since high column density systems that dominate \omg , would be
theoretically most likely to have strong dust obscuration.  Here and
throughout the paper we will use \omg\ to mean \odla , the neutral gas
contained in systems defined as being \dlas , i.e. with an \ion{H}{1}
surface density $\mnhi \geq 2 \times 10^{20} \cm{-2}$.   For a
detailed discussion of the rationale behind this choice, see PHW05.

Because radio observations are insensitive to the presence of dust, a
radio-selected sample of quasars does not suffer from this dust
obscuration selection effect.  Therefore, a radio-selected survey is a
check on the possible introduction of biases in the magnitude-limited,
optical surveys.  One previous survey, the Complete Optical and Radio
Absorption Line System (CORALS) survey by \cite{ellison01}, attempted
to answer the question of the importance of dust obscuration by
selecting quasars from the Parkes quarter-Jansky Flat-spectrum Radio
Survey \citep{jackson02}, and then following up with optical
observations of the selected quasars, regardless of magnitude, to
search for the presence of \dlas .  Ellison et al.\ found a slightly
higher incidence of \dlas\ than that found by optical surveys.  From
their measurement of \omg , Ellison et al.\ concluded that the effects
of dust could be suppressing the magnitude-limited value of \omg\ by
no more than a factor of two.

The radio-selected surveys were motivated by several studies which
indicated that dust obscuration would significantly bias the results
of \omg\ and other quantities derived from optical surveys.
\cite{fall93} constructed models to predict the possibility of dust
obscuration of quasars and found that between 10 and $70 \%$ of
quasars at $z = 3$ could be obscured, resulting in an underestimate of
\omg\ by the optical surveys.
Recently, \cite{wild05} report on a survey of SDSS
quasars for \ion{Ca}{2} absorption-line systems with $0.84 < z_{abs} < 1.3$.  
Using \ion{Ca}{2} along with Fe and Mg lines, they claim that most of
their sample likely contains \dlas\ and a significant amount of
reddening.  By modeling the reddening of these systems they make an
estimate that they are  missing $\approx 40\%$ of \ion{Ca}{2} systems from
the SDSS due to dust obscuration, which they compare favorably to the
upper limit of the CORALS survey.

On the other hand, the SDSS-DR3 survey which found 525 \dlas\ (PHW05),
indicates that a dust bias, if present, is not an important effect.
Prochaska et al. examine their results as a function of quasar
magnitude, separating out the brightest 33\% and the faintest 33\% of
the sample in each of four redshift bins.  While there is not a
statistically significant difference in the line density, they measure
$40\%$ higher \omg\ values towards brighter quasars at the $95\%$ c.l.
Since this is the opposite effect of what the dust bias would naively
imply  (a dust bias may imply that $\momg^{bright} < \momg^{faint}$,
since the brightest observed quasars should have less foreground dust
obscuration which implies a smaller \nhi\ value and hence smaller
\omg), the SDSS-DR3 results with a statistically significant higher
value of \omg\ towards brighter quasars point to the conclusion that
dust obscuration is not an important effect.  When \cite{murphy04}
examined the results of the SDSS-DR2, a sample including 70 \dlas\,
they found no evidence for reddening.  After examining the much larger
SDSS-DR3 \cite{murphy05} find evidence for reddening, but at a very
low level -- the implied dust to gas ratio is less than 0.02, where
dust to gas ratio is defined relative to that of the Milky Way (see
equation 7 of \cite{wolfe03}).  However, with all of these studies it
is important to remember that optical samples cannot measure dust bias
in objects so heavily extincted that they are missing from the
samples.  As larger optical surveys for \dlas\ become feasible, due to
surveys like the SDSS, and the statistical uncertainties become
$<5\%$, potential causes of systematic uncertainties, such as dust
obscuration, must be fully understood.

In this paper we will present the results of a radio-selected quasar
survey that was undertaken by our group.  This UCSD sample is
approximately the same size as the CORALS sample and  presents a
comparable assessment of dust obscuration.  We will analyze the
combined results, attempt to assess the \ion{H}{1} column density distribution
function, $\mfnhi$, and show that our results for \omg\ do not differ
in a statistically significant way from the results of
optically-selected surveys, therefore suggesting that dust obscuration
is most likely not a major problem affecting optically-selected quasar
samples for \dlas .

The organization of this paper is as follows:  In $\S$~\ref{sec:ucsd}
we describe the UCSD quasar sample, DLA identification method and
analysis processes.  In $\S$~\ref{sec:dla_stats} we review the
standard \dla\ statistical analysis methods.  We discuss the
results of the UCSD, CORALS and combined samples in
$\S$~\ref{sec:results}, as well our process for dealing with the Empty
Fields.  And finally, in $\S$~\ref{sec:analysis} we will compare our
results with the most recent optical surveys.

Throughout the paper we will use the following cosmological parameters \citep{bennett03}:  $\Omega_\Lambda =
0.7, \Omega_m=0.3, H_0=70$\kms\,Mpc$^{-1}$.

\section{THE UCSD SAMPLE}
\label{sec:ucsd}


The UCSD sample consists of 68 objects selected from the 411 sources
that comprise the complete Caltech-Jodrell Bank radio catalogs,
including the Pearson-Readhead sample (PR), the Caltech-Jodrell Bank
VLBI Surveys 1 and 2 (CJ1 and CJ2), and the Caltech-Jodrell Bank
Flat-spectrum sample completion.  While each sample was selected by a
progressively lower flux density threshold, the sources for each
sample were all selected to have declination (B1950) $\delta \geq
35\degr$ and Galactic latitude $\mid b\mid$ $\geq 10\degr.$  The PR
sample includes 65 objects with flux density at 6 cm (4859 MHz),
$S_{6cm} \geq 1.3$ Jy \citep{pearson88}.  The CJ1 sample includes 135
sources with flux density at 6 cm, 1.3Jy$ \geq S_{6cm} \geq $0.7Jy
\citep{polatidis95, thakkar95, xu95}, and the CJ2 sample consists of
193 mostly flat-spectrum objects with a flux density $ S_{6cm} \geq
$350 mJy \citep{taylor94, henstock95}.  The CJF
\citep[CJF;][]{taylor96} is a compilation of the flat-spectrum radio
sources (spectral index flatter than $\alpha ^{4850 MHz}_{1400 MHz}
\geq -0.5$) from the previous three surveys, plus and additional 18
sources for completion.

An optical campaign to determine the type of source, magnitude, and
redshift of the radio catalog objects followed, the results of which
were compiled in the CJ catalogs.  The object optical identification
and determination of the R magnitude was done by automated scanning of
the POSS plates or by eye.  Redshifts were primarily taken from
\cite{vernon93} and from \cite{henstock97}.

From this large radio sample, our selection criterion included all
objects identified as quasars with $z_{em}>2.0$ regardless of optical
magnitude.  The $z_{em} > 2.0$ cutoff was chosen to ensure sufficient
spectral coverage to search for \dlas\ at wavelengths redward of the
atmospheric cutoff at $\approx$ 3200\AA.  We also included in our
sample of 68 objects all 14 sources designated as optical empty fields
in order to be sure that we were not artificially selecting objects
brighter than an arbitrary optical magnitude.  And finally, we
included the 5 sources for which there was a tentative optical
identification, but no redshift information.


\subsection{Observations and Analysis}

Observations of most quasar candidate objects were first carried out
at Palomar, with follow-up done at Keck for faint or ``Empty Field''
(EF) objects (see Table~\ref{tab:observations}).  The majority
of our spectra have better than 6 \AA\ FWHM.  Five of the 68
objects in our sample were previously observed
at moderate spectral resolution and the data or results for these
objects were taken from the literature.  These included quasars
Q0014+813, Q0201+365, Q0636+680, Q0642+449 which were observed by
\cite{sargent89}, and quasar Q1124+571 which  was taken from
\cite{lzwt91}.  The \nhi\ measurement for the \dla\ towards Q0201+365 was
taken from \cite{lu93}.

\begin{table}[ht]\scriptsize
\begin{center}
\caption{{\sc DETAILS OF OBSERVATIONS\label{tab:observations}}}
\begin{tabular}{lccc}
\tableline
\tableline
Telescope &Date&Resolution&No. of\\
& (No. nights) & (at 4000 \AA ) & quasars observed\\
\tableline
Palomar&Nov 1995 (2)&4-6 \AA&17\\
Palomar&Apr 1996 (1)&4-6 \AA&8\\
Palomar&May 1996 (2)&4-6 \AA&9\\
Palomar&Dec 1996 (2)&4-6 \AA&16\\
Palomar&June 1997 (1)&4-6 \AA&7\\
Keck LRIS&Nov 2001 (1)&4-6 \AA&3\\
Keck ESI&Apr 2002 (1)&0.5 \AA&14\\
Keck ESI&Aug 2002 (1)&0.5 \AA&7\\
Keck LRIS&Dec 2002 (1)&11 \AA&1\\
Keck LRIS&May 2003 (1)&3-5 \AA&14\\
\tableline
\end{tabular}
\end{center}
\end{table}

Initial observations of most other targets were made with the 200-inch
Hale telescope of the Palomar Observatory.  Observations were made
with the Double Spectrograph and utilized gratings that were 300 lines
mm$^{-1}$ in the blue and the 315 lines mm$^{-1}$ in the red,
resulting in 4-6 \AA\ resolution using the 1\arcsec slit, and
$\approx$10 \AA\ resolution using the 2\arcsec slit when conditions
were poor.  All data were reduced using standard IRAF reduction
packages.

Follow up observations of empty fields and faint quasars were done at
Keck with the Echellette Spectrograph and Imager
\citep[ESI,][]{sheinis02} and the Low Resolution Imaging Spectrometer
\citep[LRIS,][]{oke95}.  With LRIS, the slit size was generally
0.7\arcsec.  The ESI observations utilized the 1.0\arcsec slit in LowD
mode and the 0.5\arcsec slit in echellette mode. 

Fifteen of the original 68 objects were excluded from the final
statistical data set for the following reasons:  Seven were determined
to be galaxies, stars, or low redshift quasars, and the remaining
eight were deemed empty fields (EFs).  Table~\ref{tab:ucsd} contains
the details of the final 53 quasars used in the UCSD statistics, while
Table~\ref{tab:empty} and Table~\ref{tab:discard} contain a summary of
the empty fields and the discarded objects respectively.

\begin{table*}
\begin{center}
\caption{UCSD SURVEY SAMPLE\label{tab:ucsd}}
\begin{tabular}{lccccllll}
\tableline
\tableline
Quasar &$z_{em}$ & R mag & $\mnhi \times 10^{20} \cm{-2}$ & 
$z_{abs}$ & $z_{min}$&$z_{max}$
&6cm Flux (Jy)&Survey\tablenotemark{d}\\
\tableline
Q0014+813&3.384&15.9&...&...&1.591&3.340&0.551&CJ2\\
Q0153+744&2.338&16.0&...&...&1.568&2.305&1.794&PR\\
Q0201+365&2.912&17.5&$2.4$\tablenotemark{a}&2.461&1.632&2.873&0.349&CJ2\\
Q0212+735&2.367&19.0&...&...&1.742&2.333&2.444&PR\\
Q0604+728&3.53&20.3&...&...&2.651&3.485&0.654&CJ2\\
Q0609+607&2.710&19.0&...&...&1.650&2.673&1.059&CJ2\\
Q0620+389&3.470&20.0&...&...&1.842&3.425&0.87&CJ1\\
Q0627+532&2.200&18.5&...&...&1.619&2.168&0.485&CJ2\\
Q0636+680&3.174&16.2&...&...&1.591&3.132&0.499&CJ2\\
Q0642+449&3.406&18.5&...&...&1.591&3.362&0.78&CJ1\\
Q0727+409&2.501&17.0&...&...&1.578&2.466&0.468&CJ2\\
Q0738+491&2.318&21.0&...&...&1.928&2.285&0.352&CJ2\\
Q0749+426&3.590&18.1&...&...&2.118&3.544&0.461&CJ2\\
Q0800+618&3.04&19.6&$3.16\pm 0.15$&$2.9603\pm0.0017$&2.234&3.000&0.981&CJ2\\
Q0803+452&2.102&19.6&...&...&1.594&2.071&0.414&CJ2\\
Q0824+355&2.249&19.7&$2.0$\tablenotemark{b}&2.2433&1.655&2.217&0.746&CJ2\\
Q0833+585&2.101&18.0&...&...&1.546&2.070&1.11&CJ1\\
Q0836+710&2.180&16.5&...&...&1.507&2.148&2.423&PR\\
Q0902+490&2.690&17.2&...&...&1.550&2.653&0.547&CJ2\\
Q0917+449&2.180&19.0&...&...&1.578&2.148&0.80&CJ1\\
Q0930+493&2.590&18.4&...&...&1.666&2.554&0.574&CJ2\\
Q1014+615&2.800&18.1&$2.5$\tablenotemark{b}&2.7670&2.263&2.757&0.631&CJ2\\
Q1053+704&2.492&18.5&...&...&1.801&2.457&0.71&CJ1\\
Q1124+571&2.890&18.0&...&...&1.796&2.851&0.597&CJ2\\
Q1144+542&2.201&20.5&...&...&1.632&2.169&0.88&CJ1\\
Q1155+486&2.028&19.9&...&...&1.632&1.998&0.445&CJ2\\
Q1214+588&2.547&19.5&...&...&1.632&2.512&0.307&CJ2\\
Q1239+376&3.818&19.5&$2.0\pm 0.15$&$3.4082$\tablenotemark{c}&2.344&3.770&0.446&CJ2\\
Q1325+436&2.073&18.5&...&...&1.549&2.042&0.533&CJ2\\
Q1333+459&2.449&18.5&...&...&1.612&2.414&0.76&CJ1\\
Q1337+637&2.558&18.5&...&...&1.550&2.522&0.431&CJ2\\
Q1413+373&2.360&17.3&...&...&1.607&2.326&0.383&CJ2\\
Q1421+482&2.220&18.9&...&...&1.549&2.188&0.536&CJ2\\
Q1427+543&2.980&20.7&...&...&2.331&2.940&0.718&CJ2\\
Q1435+638&2.068&15.0&...&...&1.591&2.037&1.24&CJ1\\
Q1526+670&3.020&17.1&...&...&1.977&2.980&0.417&CJ2\\
Q1547+507&2.169&18.5&...&...&1.582&2.137&0.74&CJ1\\
Q1602+576&2.858&16.8&...&...&1.630&2.819&0.351&CJ2\\
Q1624+416&2.550&22.0&...&...&1.732&2.515&1.632&PR\\
Q1645+635&2.380&19.4&$3.55\pm 0.15$&$2.1253\pm 0.0003$&1.536&2.346&0.444&CJ2\\
Q1745+624&3.886&18.3&...&...&3.085&3.837&0.580&CJ2\\
Q1755+578&2.110&18.6&$25.1\pm 0.15$&$1.9698\pm0.0009$&1.630&2.079&0.455&CJ2\\
Q1758+388&2.092&17.8&...&...&1.512&2.061&0.92&CJ1\\
Q1834+612&2.274&17.6&...&...&1.599&2.241&0.590&CJ2\\
Q1839+389&3.094&19.5&$5.0\pm 0.15$&$2.7746\pm0.0009$&1.911&3.053&0.476&CJ2\\
Q1850+402&2.120&17.9&$20.0\pm 0.25$&$1.9888\pm0.0058$&1.669&2.089&0.535&CJ2\\
Q2015+657&2.845&19.1&...&...&2.734&2.807&0.500&CJ2\\
Q2017+745&2.187&17.9&...&...&1.602&2.155&0.500&CJ2\\
Q2136+824&2.350&18.9&...&...&2.002&2.317&0.509&CJ2\\
Q2255+416&2.150&20.9&...&...&2.119&2.119&0.99&CJ1\\
Q2259+371&2.228&20.4&...&...&1.632&2.196&0.406&CJ2\\
Q2310+385&2.181&17.5&...&...&1.630&2.149&0.484&CJ2\\
Q2356+385&2.704&18.6&...&...&1.771&2.666&0.449&CJ2\\
\tableline
\end{tabular}
\end{center}
\tablenotetext{a}{\nhi\ value taken from \cite{lu93}.}
\tablenotetext{b}{Associated systems: $z_{abs}$ within $\approx$ 3,000 km s$^{-1}$ of $z_{em}$.}
\tablenotetext{c}{Weak metals, therefore best fit determined by eye.}
\tablenotetext{d}{CJ1, CJ2 = Caltech Jodrell Bank, PR = Pearson-Readhead}
\end{table*} 

\begin{table*}
\begin{center}
\caption{EMPTY OR EXTENDED FIELDS\label{tab:empty}}
\begin{tabular}{lllcccccc}
\tableline
\tableline
Object &RA(J2000)&Dec(J2000) &Exp. &R$_{lim}$$^{a}$&
Morphology &R mag$^{b}$ &6cm Flux&Survey\\
&&&time (s)&3$\sigma$&&&(Jy)&\\
\tableline
0102+480&01 05 49.93&$+$48 19 03.19&294&26.1&no detection&...&1.080&CJ1\\
0633+596&06 38 02.87&$+$59 33 22.21&500&26.4&possibly extended&25.8 $\pm 0.7$&0.482&CJ2\\
0718+793&07 26 11.74&$+$79 11 31.0&500&26.1&extended, $\approx$7$\arcsec \times$3$\arcsec$&24.5 $\pm 0.4$&0.467$^{c}$&CJ2\\
1107+607&11 10 13.09&$+$60 28 42.57&600&26.5&extended, $\approx$4$\arcsec \times$2.5$\arcsec$&25.8 $\pm 0.8$&0.400&CJ2\\
1205+544&12 08 27.50&$+$54 13 19.53&600&26.5&possibly extended&24.6 $\pm0.3$&0.397&CJ2\\
1312+533&13 14 43.83&$+$53 06 27.73&600&26.5&extended, $\approx$3$\arcsec \times$3$\arcsec$&25.4 $\pm 0.6$&0.433&CJ2\\
1828+399&18 29 56.52&$+$39 57 34.69&900&26.9&no detection&...&0.353&CJ2\\
2054+611$^{d}$&20 55 38.84&$+$61 22 00.64&900&26.7&possibly extended&...&0.414&CJ2\\
\tableline
\end{tabular}
\end{center}
\tablenotetext{a}{~Limiting magnitude per seeing element above sky background}
\tablenotetext{b}{R mag estimation of extended smudge}
\tablenotetext{c}{@1.4GHz}
\tablenotetext{d}{Uncertain identification, either $z$ = 1.588, 3.0, 3.3}
\end{table*}

\begin{table}[ht]\scriptsize
\begin{center}
\caption{{\sc DISCARDED OBJECTS\label{tab:discard}}}
\begin{tabular}{llc}
\tableline
\tableline
Object &Reason for Discard &Survey\\
\tableline
0843+575&galaxy&CJ2\\
1125+596&quasar at $z_{em}$ = 1.78&CJ2\\
1308+471&galaxy&CJ2\\
1436+763&star&CJ2\\
1809+568&No significant emission feature&CJ2\\
2238+410&Spectrum dubious&CJ2\\
2319+444&quasar at $z_{em}$ = 1.24&CJ2\\
\tableline
\end{tabular}
\end{center}
\end{table}

\subsection{Damped \lya\ systems}

The Palomar data were reduced using standard IRAF packages, while the
Keck data were reduced using IDL reduction 
software\footnote{http://www.ucolick.org/$\sim$xavier/IDL}.  The reduced
quasar spectra were continuum fitted and normalized and then analyzed
to find regions in which the restframe equivalent width of an
absorption feature was $\geq 5$ \AA\ and located in a region of good
signal to noise.  The equivalent width of the spectrum was calculated
and an equivalent width array was then analyzed for regions that were
greater than the 5 \AA\ restframe cutoff, as explained by \cite{wolfe95}.
We searched all regions of the spectrum blueward of \lya\ emission,
beginning with the lowest wavelength at which the error was below the
restframe equivalent width threshold of $\geq 5$ \AA\ at the 5$\sigma$
level.  All candidate detections were then inspected by eye to
determine if they were indeed \dlas .  False detections were usually
quite obvious to exclude as blended lines, Ly$\beta$, etc.

Nine \dlas\ were found, two of which, towards quasars Q0824+355 and
Q1014+615, were within 3,000\, \kms of the \lya\ emission peak, and
therefore considered ``associated''.  Following the standard practice,
we exclude these ``associated'' \dlas\ from the sample in order to
insure that we are not detecting objects that are physically
associated with the quasar.  Discarding these two leaves a final seven
\dlas\ to be included in the UCSD sample.

\subsubsection{\lya\ Fits}

The \dla\ systems were fitted with Voigt profiles using the IDL
tool\footnote{http://www.ucolick.org/$\sim$xavier/IDL} 
{\it x\_fitdla} which allows the user to interactively modify the
Voigt profile and continuum placement.  In all but one case, that of
Q1239+376, the \dla\ redshift was constrained by the corresponding
metal lines with errors as given in Table~\ref{tab:ucsd}.  In the case
of Q1239+376, the metal lines were too weak for use in
constraining the \dla \ redshift and we instead determined the best
fit interactively by eye using {\it x\_fitdla}.  This
method results in larger uncertainties for $z_{abs}$ and \nhi.

\begin{figure}[ht]
\begin{center}
\includegraphics[height=3.6in]{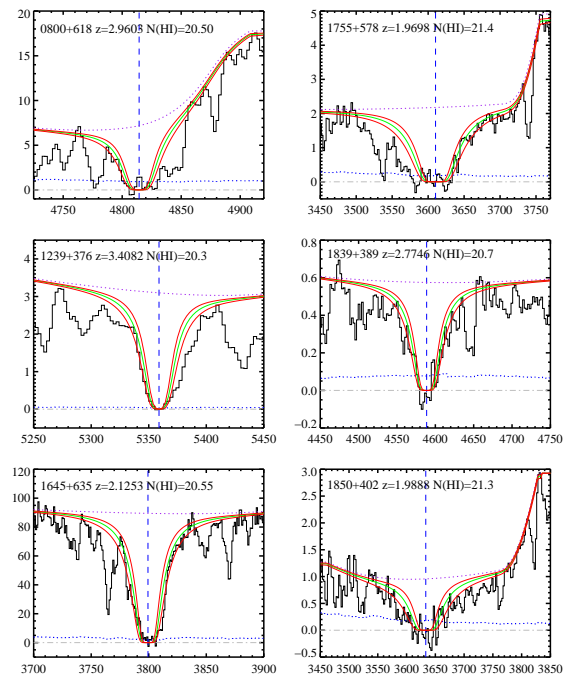}
\caption{Voigt profile \dla\ fits for the six new \dla\ systems
in the UCSD sample.  The best Voigt profile is indicated by the green
curve, surrounded by profiles with \nhi\ displacements $\pm$ 0.15\,dex
(or $\pm$ 0.25\,dex) in red.  The continuum placement can be seen as
the purple dotted line, while the error array is represented by the
blue dotted line.
}
\label{fig:dla}
\end{center}
\end{figure}

For most of our sample \dlas\, a conservative estimate of the
uncertainty in \nhi\ is 0.15\,dex.  However in one case, that of
Q1850+402, where the Voigt profile proved difficult to fit, we report
an uncertainty of 0.25\,dex in \nhi .  Figure~\ref{fig:dla} shows
Voigt profile fits for each \dla , except for the $z_{abs}= 2.461$
\dla\ towards Q0201+365, which is in the existing literature
\citep{sargent89}.  Now we will give brief details on each \dla\
system.

{\bf Q0201+365:}  \nhi\ fit taken from \cite{lu93} and discussion
therein.

{\bf Q0800+618:}  The difficulty in estimating the continuum placement in
such close proximity to the \lya\ emission peak made this \dla\ system
somewhat difficult to fit.  There is also a possibility of some blended
absorption.

{\bf Q1239+376:}  High signal to noise and good placement.  The only
problem with this fit was some blending on the red side.

{\bf Q1645+635:}  High signal to noise and lack of blending
resulted in a good Voigt profile fit to this \dla\ profile.

{\bf Q1755+578:}  Close proximity to \lya\ emission peak and blending on the
blueward side made the fit difficult.

{\bf Q1839+389:}   Straightforward fit and good continuum.

{\bf Q1850+402:}  Close proximity to \lya\ emission peak, lower signal
to noise and blending made this a more difficult fit, yielding an
increased error margin on the Voigt profile of 0.25 dex.

\section{\dla\ statistics}
\label{sec:dla_stats}

Our goal of determining the impact of dust obscuration in surveys of
\dlas\ requires that we be able to compare our radio-selected survey
to the results of optically-selected surveys.  We will now introduce
some of the standard statistical quantities used to describe and
quantify surveys of \dlas .

\subsection{$\Delta z$, $\gz\,$ and $n(z)$}

The redshift path, $\Delta z$, is defined as the total redshift
interval along which a damped \lya\ feature with rest frame equivalent
width exceeding 5\AA\ would be detected at more than $5 \sigma$
significance.  It is defined as follows,

\begin{equation}
\Delta z = \sum_{i=1}^{n} (z_i^{max} - z_i^{min})
\end{equation}

\noindent where the summation is over the $n$ quasars in the survey,
  $z_{min}$ is determined to be the lowest spectral wavelength with
  good signal-to-noise, and $z_{max}$ is the redshift corresponding to
  the maximum spectral wavelength included in the search.  We define
  $z_{max}$ by,

\begin{equation}
z_{max} \equiv z_{qso} - (1 + z_{qso})/100\perd
\end{equation}

\noindent  This corresponds to 3000\,\kms blueward of the \lya\ emission
feature.  This cutoff ensures that a damped \lya\ system is not
physically associated with the quasar.

The redshift path density, \gz, gives an idea of the statistical
significance as a function of redshift of a survey for \dlas .  It is
defined as the number of quasars with sightlines containing a particular
redshift interval \citep{lzwt91}.  Specifically,

\begin{equation}
g(z) = \sum_{i=1}^{n} H(z_i^{max} - z)H(z - z_i^{min})\cmma
\end{equation}

\noindent where H is the Heaviside step function, the sum is over $n$
quasars \citep{lzwt91} and, 

\begin{equation}
\Delta z = \int g(z) dz \cmma
\end{equation}

\noindent where the integral is over all $z$ paths in the survey.  The
\dla\ number density, $n(z)$, is simply the number of \dlas\
per unit redshift,

\begin{equation}
n(z) = \frac{m}{\Delta z} \cmma
\end{equation}

\noindent where $m$ is the number of \dlas .

\subsection{\fnhi\ : The \ion{H}{1} Frequency Distribution Function}

Following the direction of previous works such as
\cite{lzwt91} and PHW05, we can define a neutral hydrogen frequency
distribution function that describes the number of \dlas\ in a range
of column densities, $(N, N+dN)$, and a range of absorption distances,
$(X, X+dX)$,

\begin{equation}
f_{\rm HI}(N,X) dN dX \cmma
\end{equation}

\noindent where the absorption distance, $\Delta X$, is defined as follows:

\begin{equation}
\Delta X = \int dX \equiv \int \frac{H_0}{H(z)} (1+z)^2 dz
\end{equation}

\noindent where $H_0$ is Hubble's constant.

\subsection{$\ldla(X)$ : The Damped \lya\ Line Density}

The zeroth moment of the \ion{H}{1} frequency distribution function is
known as  the line-density of \dlas\, $\ldla (X)$.  The line-density
represents the number of systems per unit absorption distance and is
defined as:

\begin{equation}
\ldla(X) = \intl_{N_t}^\infty \mfnhi dN \perd
\end{equation}

\noindent  As described in PHW05, the line-density is
related to the covering fraction of \dlas\ on the sky.  This relationship is
apparent if we describe the frequency distribution function in terms
of an average cross-section $A(X)$, and the comoving number density of \dlas\,
$n_{DLA}(X)$:

\begin{equation}
\mfnhi \equiv (c/H_0) n_{DLA}(X) A(X) \perd
\end{equation} 

\noindent \citep[see][for details]{wolfe05}. 

\subsection{\omg: The Cosmological Neutral Gas Mass Density}
\label{sec:omg}

An important parameter
in describing any quasar survey for \dlas\ is the first moment of the
\ion{H}{1} frequency distribution function, the neutral gas mass density,
\omg.  It is believed  that \omg\ is closely related to the amount of
neutral hydrogen available for star formation and hence, places an
important tracer on the history of star formation in the Universe.
Acquiring this parameter through surveys for \dlas\ is an important
constraint on the neutral gas reservoir available for star formation in
the early ($z > 2$) Universe.  \omg\ is defined as follows:

\begin{equation}
\Omega_g(X) \equiv \frac{\mu m_H H_0}{c \rho_c} 
  \intl_{N_{min}}^{N_{max}} N \mfnhi dN
\label{eqn:omg}
\end{equation}

\noindent where $\mu$ is the mean molecular mass of the gas (taken to
be 1.3), m$_H$ is the mass of the hydrogen atom, $\rho_c$ is the critical mass
density, $\mfnhi$ is the frequency distribution function of
neutral gas found in \dlas\, and the integration is from $N_{min} = \Nth$ to
$N_{max} = \infty$.
We follow previous works, i.e. \cite{lzwt91}, and replace this
frequency distribution function by its evaluation in the discrete
limit as follows,

\begin{equation}
\Omega_g = \frac{\mu m_H H_0}{c \rho_c} \frac{\Sigma \N{HI}}{\Delta X} \;\; ,
\label{eqn:omgdisc}
\end{equation}

\noindent where the sum is performed over the \nhi\ measurements of
the \dla\ systems in a given redshift interval with total pathlength
$\Delta X$.   As emphasized by PHW05 and discussed below, 
equation~\ref{eqn:omgdisc} only provides an accurate evaluation
of equation~\ref{eqn:omg} if the survey is sufficiently large that
the observed \fnhi\ distribution becomes steeper than $N^{-2}$ at large \nhi .
If this is not the case, equation~\ref{eqn:omgdisc} provides only a 
lower limit to \omg.

\section{Results}
\label{sec:results}

We will now describe the results of the UCSD radio-selected survey,
the CORALS radio-selected survey, and the combination of these two
surveys, which we will refer to as the combined sample.  Details of the
results of each survey are listed in Table~\ref{tab:results}.

\subsection{UCSD Survey Results}

\begin{figure}[ht]
\begin{center}
\includegraphics[height=3.6in,angle=90]{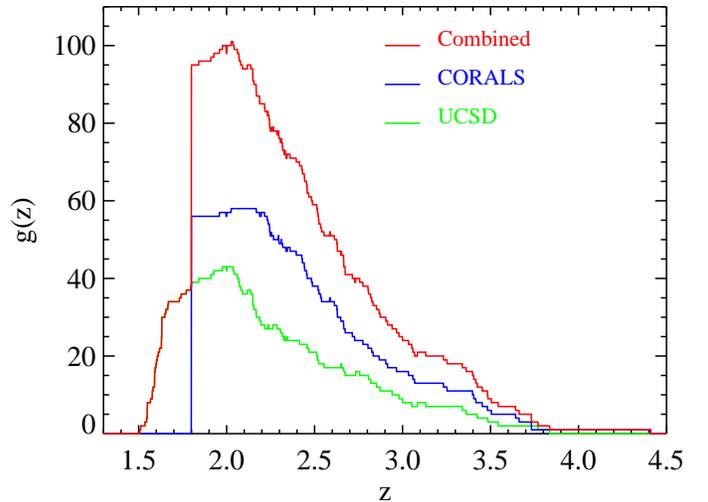}
\caption{Redshift sensitivity function $g(z)$ as a function of redshift for
the UCSD survey, the CORALS survey, and the combined sample.
}
\label{fig:gzsmpl}
\end{center}
\end{figure}

The UCSD sample consists of 7 \dlas\ in 53 quasars of $z_{em} \geq
2.0$ with a total redshift path of $\Delta z = 41.15$.  This resulted
in a number of \dlas\ per unit redshift, $n(z) =
0.17^{+0.08}_{-0.07}$, where the error bars are the standard $1\sigma$
Poissonian errors using Gehrels' tables for small number statistics
\citep{gehrels86}.  Figure~\ref{fig:gzsmpl} presents \gz\ versus $z$
for the UCSD sample in green.  The line density of \dlas\ over the
cosmological redshift path of $\Delta X = 130.43$ resulted in
$\ldla(X) = 0.05^{+0.03}_{-0.02}$, while the mass density of neutral
gas is $\momg = 0.84^{+0.43}_{-0.45}\times 10^{-3}$. 
While we report $\momg$ as a detection, it is strictly speaking, a lower
limit because we do not measure \fnhi\ to fall off faster than $N^{-2}$
(see $\S$~\ref{sec:combo}).

\subsection{CORALS Survey Results}
\label{sec:corals}

The Complete Optical and Radio Absorption Line Survey (CORALS)
\citep{ellison01} was the first attempt to utilize a radio-selected
quasar survey as a basis for a search for \dlas .  They selected
quasars from the complete Parkes quarter-Jansky flat spectrum sample
\citep{jackson02}, comprised of 878 radio sources with spectral index
$\alpha ^{5 GHz}_{2.7 GHz} > -0.4$ and declinations between
$+2.5\degr$ and $-80\degr$.
\cite{ellison01} limited their data set to 66 $z_{em} \geq 2.2$
quasars in which they found 22 \dlas .  Three of these \dlas\ were
classified as ``associated'' and dropped from the final sample.  Two
more \dlas\ were excluded because they fell outside of the range $1.8
\leq z_{abs} \leq 3.5$, that \cite{ellison01} set for their
statistical sample, citing that there appears to be little evolution
of \omg\ in this range.  Since the UCSD sample did not have this
$z_{abs}^{max}$ cutoff, we included all 19 CORALS \dlas\ in the
statistics for the combined sample.

Detailed results of the CORALS survey, including the two \dlas\ in
sightlines to  quasars with $z_{abs} > 3.5$, are listed in
Table~\ref{tab:results}. Statistics for the number density and neutral
gas mass density in the CORALS survey resulted in their conclusion
that previous,  magnitude limited surveys could have underestimated
these values by as much as a factor of two.  The plot of \gz\ versus
$z$ for the CORALS survey is shown in blue in Figure~\ref{fig:gzsmpl}.
Over a total redshift interval, $\Delta z = 57.16$, the number of \dla
s per unit redshift, $n(z) = 0.33^{+0.10}_{-0.07}$.  Over a
cosmological redshift path of $\Delta X = 186.68 $, the line density
of \dlas\ in the CORALS survey,
$\ldla (X) = 0.102^{+0.03}_{-0.02}$, which is double that of the UCSD
survey.  CORALS compared their neutral gas mass
density in \dlas\, $\momg = 1.37^{+0.53}_{-0.55} \times 10^{-3}$ 
with the compilation of \cite{peroux01} and \cite{rao00}, and
concluded that \omg\ derived from optically-selected surveys could be
underestimated by up to a factor of two.

However, Ellison et al. do concede the uncertainty of their conclusion
primarily  because the small survey fails to fully sample the column
density  distribution and secondly because their high value of \omg\
is dominated by two relatively high column density systems (both
incidentally in front of ``moderately bright'' quasars, $B=19.5, 20$,
which qualitatively matches the result of PHW05 that there is
an anti-correlation between quasar magnitude and \nhi).

\begin{table}[ht]\scriptsize
\begin{center}
\caption{{\sc RESULTS\label{tab:results}}}
\begin{tabular}{cccc}
\tableline
\tableline
feature&UCSD & CORALS &COMBINED\\
\tableline
No. quasars&53&66&119\\
No. \dlas\ &7&19&26\\
$\Delta$$z$&41.15&57.16&98.31\\
n($z$)&$0.17^{+0.08}_{-0.07}$&$0.33^{+0.10}_{-0.07}$&$0.26^{+0.06}_{-0.05}$\\
$\ldla(X)$&$0.05^{+0.03}_{-0.02}$&$0.10^{+0.03}_{-0.02}$&$0.08^{+0.02}_{-0.02}$\\
$\langle \mnhi \rangle$ cm$^{-2}$&$8.744\times10^{20}$&$7.532\times10^{20}$&$7.858\times10^{20}$\\
$\langle$ $z$ $\rangle$&2.53&2.50&2.51\\
$\langle$ $z$ $\rangle_{weighted}$&2.17&2.33&2.28\\
$\Sigma \mnhi$ $cm^{-2}$&$0.61\times10^{22}$&$1.43\times10^{22}$&$2.04\times10^{22}$\\
$\Delta$X&130.43&186.68&317.11\\
$\Omega_g(\times10^{-3})$&$0.84^{+0.43}_{-0.45}$&$1.37^{+0.53}_{-0.55}$&$1.15^{+0.37}_{-0.38}$\\
Error&$\pm\ 54 \%$&$\pm\ 41 \%$&$\pm\ 33 \%$\\
\tableline
\end{tabular}
\end{center}
\end{table}

\subsection{Combined Results}
\label{sec:combo}


For simplicity, we present detailed analyses for just the combined
sample, which has the greatest statistical significance.
Figure~\ref{fig:gzsmpl} presents \gz\ versus $z$ for the combined
sample in red.  The CORALS sample begins abruptly at $z = 1.8$, the
$z_{min}$ cutoff of their sample.  The UCSD sample continues down to a
$z_{min} \approx 1.51$ for some quasars.  For $z \approx 2$ the
combined sample is nearly double that of CORALS.  For higher redshift
intervals (i.e. $z$ = 3), the CORALS survey contributes roughly 2/3 of
the pathlength.
Of course, the combined sample gives the best constrained estimate of
the number density of $\nz = 0.26_{-0.05}^{+0.06}$.

\begin{figure}[ht]
\begin{center}
\includegraphics[height=3.6in,angle=90]{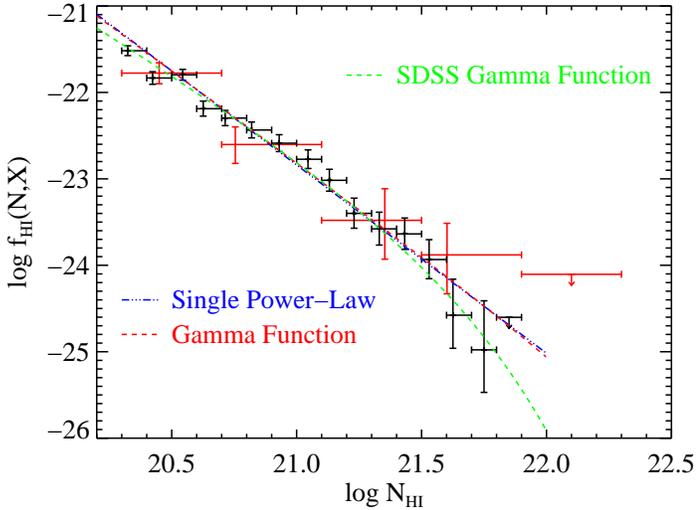}
\caption{The \ion{H}{1}  frequency distribution \fnhi\ for the 26
\dlas\ of the combined sample is plotted in red.   Overplotted are the fits of a single
power-law, the dot-dashed line in blue, and a $\Gamma$-function, the
dashed line in red.  The last bin contains the 2$\sigma$ upper limit.  Plotted in black is the \fnhi\ for the optical data from the SDSS-DR3, with the $\Gamma$-function fit in green.       
 }
\label{fig:fnall}
\end{center}
\end{figure}

The combined sample is large enough to attempt an analysis of the \ion{H}{1}
distribution function, \fnhi.  This sample spans the redshift interval $z =
[1.51, 4.4]$ with an integrated absorption pathlength $\Delta X =
317.11$ and a column density weighted mean redshift of 2.28. In
practice, we can evaluate \fnhi\ in the discrete limit and plot the
resulting \fnhi\ in \nhi\ bins of some $\Delta N$.  In
Figure~\ref{fig:fnall}, we plot in red \fnhi\ for the combined sample, in
\nhi\ bins of $\Delta N = 0.4$\,dex, calculated in the following way:

\begin{equation}
f_{\rm HI}(N,X) = \frac{m_{DLA} (N,N+\Delta N)}{\Delta X} \cmma
\end{equation}

\noindent  where $m_{DLA}$ is the number of damped \lya\ systems
within $(N, N+\Delta N)$ in the $\Delta X$ interval and the error bars
are determined by Poisson uncertainty at the 84$\%$ c.l. according to
the value of $m_{DLA}$.  Also plotted, in black, are the results of
the optically-selected SDSS-DR3 (PHW05) survey for comparison.
Following PHW05, we have overplotted the best-fit solutions of two
possible functional forms of \fnhi .  Because of the small sample size
of this survey we will attempt to fit only a single power-law and a
$\Gamma$-function.  The single power-law form is as follows:

\begin{equation}
f_{\rm HI}(N,X) = k_1 N^{\alpha_1} \cmma
\label{eqn:sngl}
\end{equation}

\noindent and the $\Gamma$-function is \citep[e.g.][]{fall93}:

\begin{equation}
f_{\rm HI}(N,X) = k_2 \ltp \frac{N}{N_\gamma} \rtp^{\alpha_2} 
\exp \ltp \frac{- N }{N_\gamma} \rtp \perd
\end{equation}

\noindent We have performed a maximum likelihood analysis to constrain
the parameters and set the constants $k_1$ and $k_2$.  A summary of
the fit parameters, along with those of the optically-selected
SDSS-DR3 survey for comparison, is given in Table~\ref{tab:fnfits}.
The best fit slope of the single power-law is $\alpha_1$ =
$-2.18^{+0.20}_{-0.26}$.  This single power-law slope can be compared
favorably with the optical SDSS-DR3 survey single power-law slope of
$\alpha_1$ = $-2.19^{+0.05}_{-0.05}$ over their entire redshift range,
$z$ = [2.2, 5.5].  This correspondence is expected because the
radio-selected survey is dominated by the low column density end which
matches that of the optical, and can be seen as a confirmation of the
two techniques.  

While the single-power law gives a good fit to the radio-selected
data,
we also attempt to fit the $\Gamma$-function for the following two
reasons: First, the single-power law is unphysical, i.e. the fit must
turn over in order for \omg\ to converge, and second, unlike the
single-power law, the $\Gamma$-function provided a satisfactory fit to
the optically-selected data.   However, unlike the optically-selected
sample, the radio-selected sample gives nearly the same fit for the
$\Gamma$-function as for the single power-law, $\alpha_2$ =
$-2.12^{+0.22}_{-0.27}$.  While we derive a formal value of the break
column density N$_\gamma$ = $22.49^{+0.29}_{-0.36}$, we interpret this
as an unrealistic extrapolation of the data.  Rather, the small size
of the radio-selected sample cannot reliably determine a break column
density, and therefore we cannot demonstrate that our \omg\ converges.

To determine if the radio-selected data rules out the
optically-selected $\Gamma$-function fit, we performed a chi-squared
test on the radio-selected data and optically-selected
$\Gamma$-function fit.  The results of the chi-squared test,
Prob$_1$($\chi^{2}$ $>$ 5.96) = 1.5$\%$ show that we can reject the
fit at the 5$\%$ level, but not at the 1$\%$ level of confidence.
While this may be evidence for modest disagreement between the two
samples, we interpret this disagreement to be primarily due to the
fact that we cannot constrain the radio-selected fit at large \nhi\ ,
i.e. the radio-selected sample does not contain enough \dlas\ to fully
sample the \ion{H}{1} distribution function.   We note that a more conservative
two-sided KS test shows agreement between the radio-selected data and
the optically-selected $\Gamma$-function fit at the 77$\%$ level.

\begin{table}[ht]\scriptsize
\begin{center}
\caption{{\sc FITS TO $f_{\rm HI}(N,X)$\label{tab:fnfits}}}
\begin{tabular}{cccc}
\tableline
\tableline
Form &Parameters& SDSS-DR3 Optical$^a$$^,$$^b$
& Combined Radio$^c$\\
\tableline
Single & $\log k_1$ &$23.36$&$22.90$\\
& $\alpha_1$ &$-2.19^{+0.05}_{-0.05}$&$-2.18^{+0.20}_{-0.26}$\\
Gamma & $\log k_2$ &$-23.52^{+0.02}_{-0.02}$&$-25.97^{+0.09}_{-0.08}$\\
& $\log N_\gamma$ &$21.48^{+0.07}_{-0.10}$&$22.49^{+0.29}_{-0.36}$\\
& $\alpha_2$ &$-1.80^{+0.06}_{-0.06}$&$-2.12^{+0.22}_{-0.27}$\\
\tableline
\end{tabular}
\end{center}
\tablenotetext{a}{\cite{pro05}}
\tablenotetext{b}{Mean absorption redshift = 3.06}
\tablenotetext{c}{Mean absorption redshift = 2.28}
\end{table}

\begin{figure}[ht]
\begin{center}
\includegraphics[height=3.6in,angle=90]{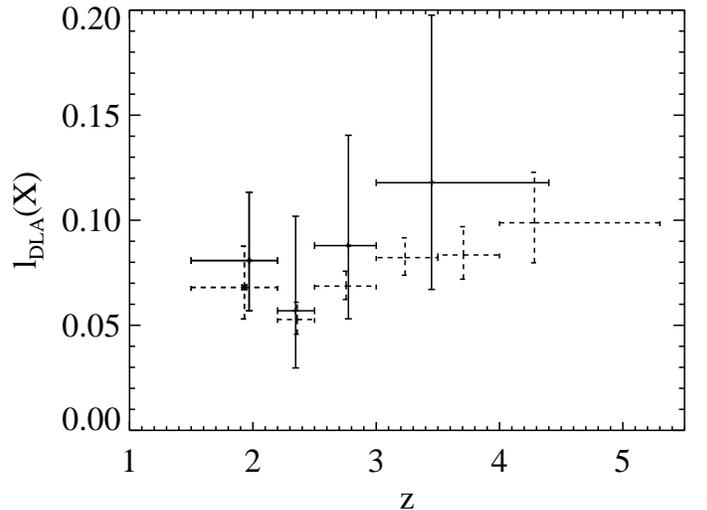}
\caption{Plot of the line density of \dla\ systems $\ldla(X)$ versus
redshift for the combined sample (solid lines).  Overplotted is the
$\ldla(X)$ for the
optical data from the SDSS-DR3 survey and the P\'eroux compilation (dashed lines).
}
\label{fig:taux}
\end{center}
\end{figure}

The line density of \dlas\ in the combined sample, taken over the
entire redshift interval, $z$ = [1.5, 4.4], is $\ldla (X) =
0.08^{+0.02}_{-0.02}$ at a median $z = 2.35$, where the errors
represent the  $1\sigma$ Poisson uncertainty in m$_{DLA}$.   In
Figure~\ref{fig:taux} we plot $\ldla (X)$ for the combined sample in
red, evaluated in the discrete limit,

\begin{equation}
\ldla(X) = \frac{m_{DLA}}{\Delta X} \perd
\end{equation}

\noindent We have grouped the data into four redshift bins, $z$ =
	  [1.5, 2.2], [2.2, 2.5], [2.5, 3.0], and [3.0, 4.4] to allow
	  for comparison with the results of PHW05.  We have
	  overplotted the results of the SDSS-DR3 optical survey in
	  black.  These points are grouped into redshift bins $z$ =
	  [2.2, 2.5], [2.5, 3.0], [3.0, 3.5], [3.5, 4.0], [4.0, 5.3].
	  The black point marked by a star in redshift bin $z$ = [1.5,
	  2.2] is a compilation of optical surveys for \dlas\ produced
	  by \cite{peroux03}.  Although the central values of the line
	  densities of the radio and optically-selected surveys are
	  different, the difference is not statistically significant.
	  Note the radio sample gives a somewhat higher line density
	  at all redshifts, and it is interesting to note that
	  beginning at $z=2.2$, the trend of increasing line density
	  with increasing redshift, $z>2.2$, is present in both
	  samples.   In fact, the central values of the radio sample
	  follow the same qualitative shape, even the unusual `dip' at
	  $z\approx 2.3$.  While it would make sense that the SDSS-DR3
	  survey, with its statistically significant numbers of
	  quasars and \dlas\, is actually detecting a physically
	  meaningful trend -- PHW05 claim the decline in $\ldla (X)$
	  is due to a decrease in \dla\ cross-section with time  --
	  the correspondence with the  combined radio sample, of
	  relatively so few objects, is likely a coincidence.
	  However, although the error bars are large, we can interpret
	  this similarity in line density evolution with the
	  statistically significant results of the SDSS-DR3 as support
	  of our results.


\begin{figure}[ht]
\begin{center}
\includegraphics[height=3.6in,angle=90]{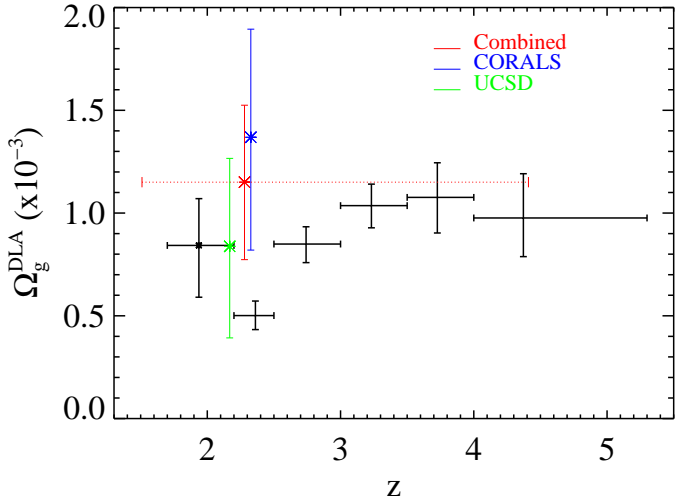}
\caption{The neutral gas mass density, $\Omega_g$, of the UCSD, CORALS and
combined samples are plotted in green, blue and red respectively.  For clarity, only the redshift bin of the combined sample is plotted, $z$$_{combined}$ = [1.51, 4.41](dotted red).  The
redshift bins of the radio samples (not plotted) are $z$$_{ucsd}$ = [1.51, 3.84], and $z$$_{corals}$ = [1.80,
4.41] and the points are plotted at the \nhi\ weighted mean redshift. 
Also plotted are the values of \omg\ as a function of redshift for the
optical SDSS-DR3 survey (black points).  The P\'eroux compilation data point at
$z<2.2$ (marked with a cross) does not include measurements from the
SDSS survey.  All error bars are 1 $\sigma$.  
}
\label{fig:omega}
\end{center}
\end{figure}

While we report a detection of the neutral gas mass density, $\momg =
1.15^{+0.37}_{-0.38} \times 10^{-3}$, our result is actually a lower
limit due to the insufficient size of the combined sample.  In
Figure~\ref{fig:omega} we plot \omg\ for each of the UCSD, CORALS and
combined samples.  Errors are calculated using a modified bootstrap
method as described by PHW05, and the values are plotted at the \nhi\
weighted mean redshift.  Also plotted, in black, are the \omg\ values
determined by the optically-selected SDSS-DR3 survey (PHW05), which
covers a redshift range $z \geq 2.2$.  Because the SDSS-DR3 sample
consists of over 500 \dlas\, the values of \omg\ are plotted in five
redshift bins: $z = [2.2, 2.5]$, [2.5, 3.0], [3.0, 3.5], [3.5, 4.0],
and [4.0, 5.5].   And finally, plotted in the bin range $z = [1.7,
2.2]$, is the compilation by \cite{peroux03}.  It is seen that the
lower limits of all of the radio-selected samples agree well with the
optically-selected data.

\subsection{Empty Fields}
\label{sec:EF}

\begin{figure}[ht]
\begin{center}
\includegraphics[height=1.3in]{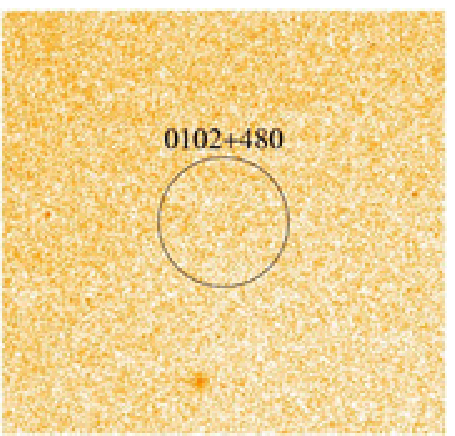}
\includegraphics[height=1.3in]{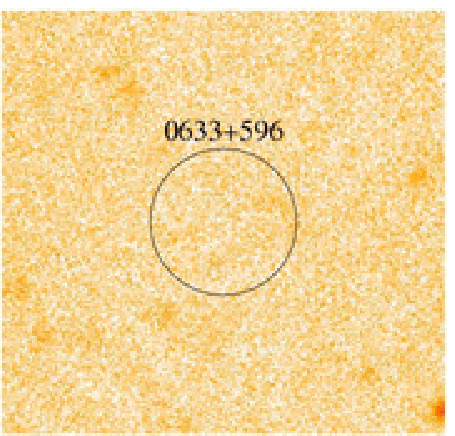}
\includegraphics[height=1.3in]{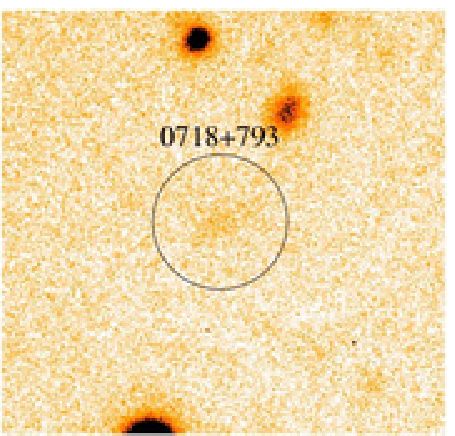}
\includegraphics[height=1.3in]{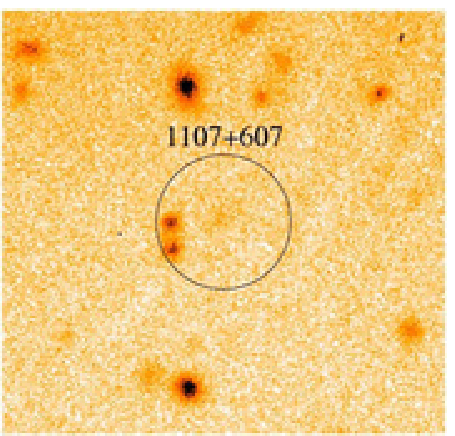}
\includegraphics[height=1.3in]{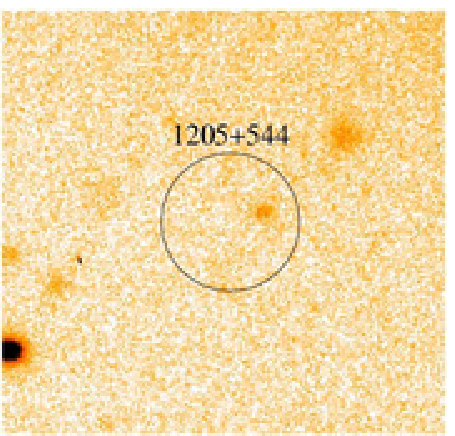}
\includegraphics[height=1.3in]{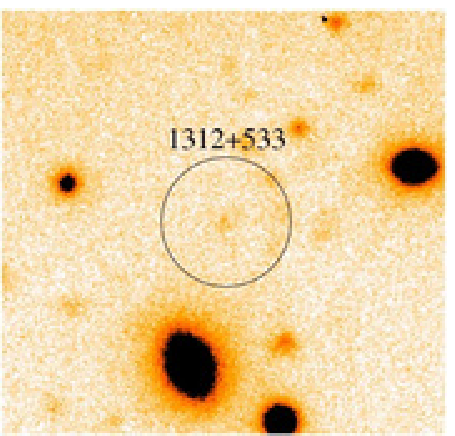}
\includegraphics[height=1.3in]{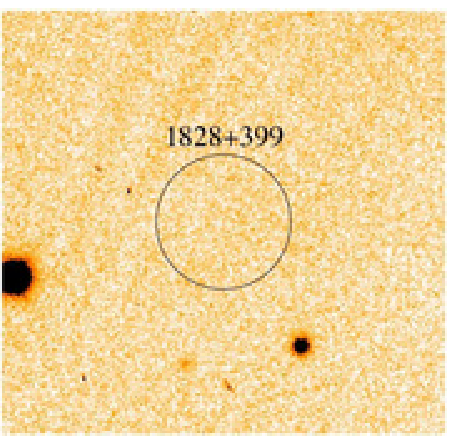}
\includegraphics[height=1.3in]{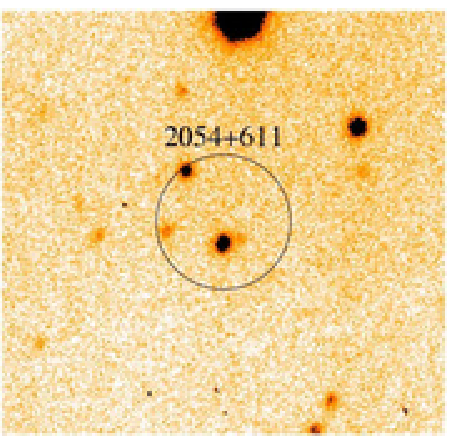}
\caption{R band images of the empty or extended fields.  Exposures times are given in Table ~\ref{tab:empty}.  The circle radius is $\approx$ 5 arcsec.  All image orientations are the standard North (up), East (left).   
}
\label{fig:empty_fields}
\end{center}
\end{figure}

The eight fields for which no optical identification of a quasar was
obtained are called the ``Empty Fields'' (EFs).  Table~\ref{tab:empty}
contains the details of each EF while Figure~\ref{fig:empty_fields}
contains Keck images of the EFs.  These eight fields were determined
to contain either nothing of significance, or merely a faint extended
smudge when imaged with Keck in the R band for $\approx$ 600 seconds.
In either case, it was not possible to obtain spectra.  While all of
the previous analyses in this paper were conducted as if these fields
did not exist, we actually must determine a method of including them
in the sample in order for our survey to be considered
complete. Assuming that the fields were truly empty, i.e. the optical
source was fainter than our magnitude limit on Keck, and that pointing
errors or some other experimental errors did not result in
radio-source misidentification, we can make two extreme, simplifying
assumptions.  On the one hand, we can assume that no \dlas\ are
present toward these optically faint quasars and calculate a lower
limit on \omg\ by including some average redshift path length for each
object, where the average pathlength is determined from the known
quasars in our survey.  On the other extreme we can assume that each
EF is actually empty because of the presence of a high column density,
dusty \dla.  We can assume that each EF contains an average to high
column-density \dla\ and estimate an upper limit on \omg.


\begin{figure}[ht]
\begin{center}
\includegraphics[height=3.6in,angle=90]{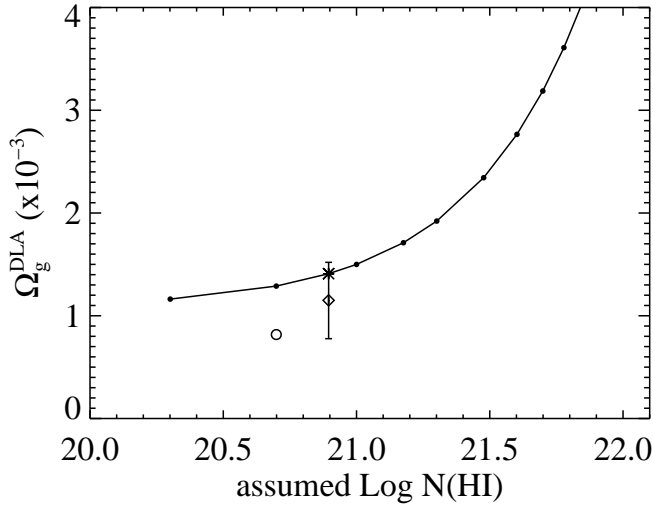}
\caption{This figure demonstrates the potential impact that the empty
fields could have on \omg , the neutral gas mass density.  We
calculate the resultant value of \omg\ assuming that each of the eight
empty fields contains a \dla\ of the stated \nhi .  The diamond with
1$\sigma$ error bars is the value of \omg\ found from the combined radio
sample, calculated by ignoring the empty fields.  The asterisk
represents the value of \omg\ if each
of the empty fields contained a \dla\ of our average $\mnhi\ = 7.86
\times 10^{20} \cm{-2}$.  For reference, the optically selected SDSS
survey value is plotted as an unfilled circle. 
 }
\label{fig:ef}
\end{center}
\end{figure}

Using the results of the combined sample, we determine the average
redshift path per quasar to be $<\Delta X> = 2.66$.  Assuming that
each of the eight EFs would contribute this amount of redshift path
gives a new total redshift path searched of $\Delta X = 338.43$.  The
lower limit on \omg , assuming that none of the EFs contained a \dla ,
is
$\momg^{lower} = 1.08 \times 10^{-3}$.  The upper limit, assuming that
each of the EFs contains a \dla\ of average column density in our
survey, $\mnhi\ = 7.86 \times 10^{20} \cm{-2}$, results in an upper
limit of
$\momg^{upper} = 1.41 \times 10^{-3}$.  While both the lower and upper
limit on \omg , derived by including the EFs, are clearly within the
error of the combined value of $\momg = 1.15^{+0.37}_{-0.38}
\times10^{-3}$, it is notable that the upper limit is only $\approx
22\%$ larger than $\momg$, i.e. even if each EF contains a dusty \dla\
of our average \nhi , the effect on \omg\ would be relatively small.

We can take the analysis one step further and allow the average value
of \nhi\ to exceed $7.86 \sci{20} \cm{-2}$.  To determine the minimum
average column density \dla\ that would affect our results we assume
that  each EF contains a quasar at our average redshift and a \dla\ of
fixed column density which we vary from the lower limit, $\mnhi = 2
\times 10^{20} \cm{-2}$, to the generally observed highest column
densities of $\mnhi \approx 1 \times 10^{22} \cm{-2}$.  We plot the
results in Figure ~\ref{fig:ef}.

As previously stated, if we assume that each EF contains a \dla\ of
average column density in our survey, $\mnhi\ = 7.86 \times 10^{20}
\cm{-2}$, we derive an $\momg = 1.41 \times 10^{-3}$, indicated in
Figure ~\ref{fig:ef} by the blue asterisk.  Compare this with the red
point and error bar, our radio-selected survey value of $\momg$,
derived by ignoring the EFs.  For reference, the optically selected
SDSS-DR3 survey, at
$\momg = 0.82^{+0.05}_{-0.05}\times10^{-3}$, is plotted in green.
 From Figure ~\ref{fig:ef} it is seen that if each EF contained a
 \dla\ of $\mnhi \approx 10^{21.2} \cm{-2}$ or larger, a relatively
 large value occurring in only $\approx 15\%$ of our sample, the impact
 of the EFs would be large enough to increase the resultant value of
 \omg\ by $\approx$ 50$\%$.

There are, however, several arguments for why these EFs are most
likely {\it not} $z_{em}>$ 2 quasars extinguished by very dusty, high
column density \dlas .  If we extrapolate the \ion{H}{1} frequency
distribution function, \fnhi\,  resulting from the
radio-selected sample, as seen in Figure ~\ref{fig:fnall}, we would
expect
not more than 1 \dla\ with $\mnhi > 10^{22} \cm{-2}$.  If two or
more high column density systems existed, the resulting $f_{\rm
HI}(N,X)$ would be unphysical assuming galaxies have declining surface
density profiles.  In this case, we would require a bimodal population
consisting of high column density, high dust-to-gas ratio systems,
such as molecular clouds, that would be missed in optical surveys.

\begin{figure}[ht]
\begin{center}
\includegraphics[height=3.6in,angle=90]{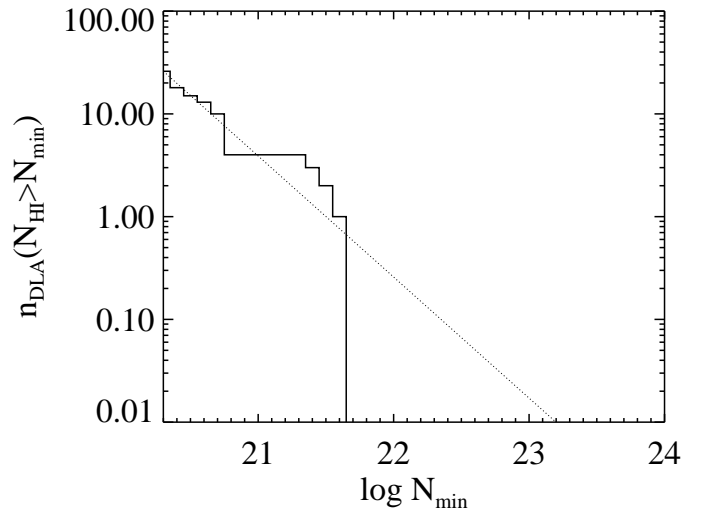}
\caption{The cumulative distribution of the number of \dlas\ with a
specific minimum $\mnhi$.  Overplotted is the single power-law
fit to the distribution function (dotted line).  It is seen that only 0.3 \dlas\ are
expected with $\mnhi\ > 10^{22} \cm{-2}$.      
}
\label{fig:cumdla}
\end{center}
\end{figure}

While we cannot rule out the existence of a bimodal population, we can
determine exactly how many high column density systems the current
radio-selected distribution function would predict.  We plot the
cumulative number of \dlas\ above a certain minimum \ion{H}{1} column density
and extrapolate using our single power-law fit.  From this plot, in
Figure ~\ref{fig:cumdla}, it is apparent that we would expect only 0.3
\dlas\ with $\mnhi > 10^{22}\cm{-2}$.
Additionally, in the case of the five fields containing faint extended
emission, we can use scaling arguments to show that if we assume this
emission is actually the resolved quasar host galaxy, then the quasar
would have to be a low redshift object and would not have been
included in our $z$$_{em} >$ 2.0 survey.  Adopting the typical high-$z$
quasar host galaxy scale length of $\approx 12$ kpc
\citep{kuhlbrodt05}, we can estimate the angular size at $z$ = 2.0 to be
$\approx$ 1.4 arcsec.  Careful inspection of the extended fields
reveals extended blobs on the order of 3 arcsec or larger, making
these low $z$ quasars that would not have been included in our survey.

In an effort to determine the nature of the EFs, we are currently
conducting an observing program on the Green Bank Telescope (GBT) to
search for 21 cm absorption along the sightlines to these EFs.  We
will carry out a redshift path search from $z$ = [0.5, 3.9] using the
frequencies of $\approx$ 300 MHz - 900 MHz.  If a high column density,
dusty system does exist along the line of sight and is blocking out
the quasar light, it is likely that we will detect it in absorption.

\section{Analysis \& Discussion}
\label{sec:analysis}

The best way of determining the significance of dust obscuration of
quasars is to compare the results of magnitude limited and
radio-selected quasar surveys for \dlas .  The primary problem with
this method has so far been the limited survey size of the
radio-selected surveys and the resultant large error bars that
preclude conclusive results.  While the UCSD survey itself was
slightly smaller than the previously published CORALS survey,
combining the two surveys in effect doubles the size.   However, the
uncertainties of the combined sample, whose size is still more than an order of
magnitude smaller than the current optical samples, are so large that
definitive statements remain elusive.

\begin{figure}[ht]
\begin{center}
\includegraphics[height=3.6in,angle=90]{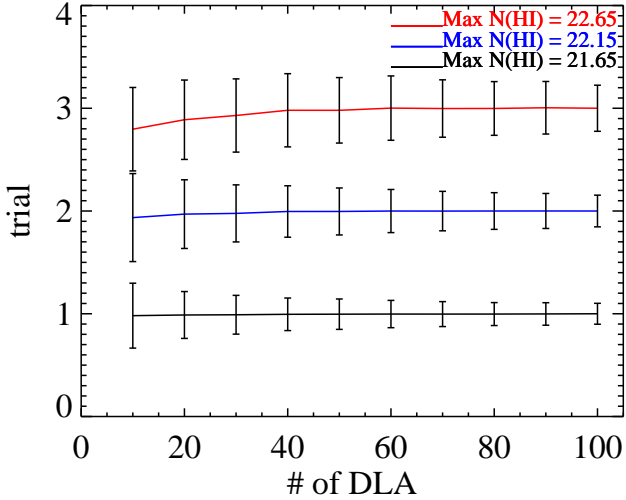}
\caption{The results of bootstrapping error on samples of the given
   number of \dlas\ with maximum column densities of $\mnhi\ =
   10^{21.65} \cm{-2}$ (to match the combined sample), the black line,
   $\mnhi\ = 10^{22.15} \cm{-2}$, the blue line, and  $\mnhi\ =
   10^{22.65} \cm{-2}$, the red line.  It is seen that a sample with
   maximum $\mnhi\ = 10^{21.65} \cm{-2}$, will give an error in the
   desired range of $\pm$10$\%$ at $\approx$ 100 \dlas . 
 }
\label{fig:smplboot}
\end{center}
\end{figure}

To make a simple estimate of the number of radio selected \dlas\
necessary to conclusively answer the question of dust bias, we
performed the following analysis.  If we desire a result with errors
of no larger than $\approx 10\%$, we can perform a bootstrap error
evaluation on a random sample of \dlas\, each time increasing the
number of \dlas\ to determine how many \dlas\ are necessary to give
the desired precision.  In Figure ~\ref{fig:smplboot} we plot the
results of our bootstrap error estimation, normalized to one and
offset for each different sample type.  The estimation was performed
on random \nhi\ samples with minimum \nhi\ = 10$^{20.3} \cm{-2}$, and
three different maximum column densities, \nhi\ = 10$^{21.65}
\cm{-2}$, to match the upper limit of the combined radio sample, \nhi\
= 10$^{22.15} \cm{-2}$, and \nhi\ = 10$^{22.65} \cm{-2}$.  It can be
seen that in a sample with a maximum column density similar to the
combined sample, the desired $\approx 10\%$ error is acheived with a
sample of $\approx$ 100 \dlas .  As the maximum column density is
increased, the error bars increase as well.  Note that the bootstrap
errors of our actual sample (number of \dlas\ = 26, error $\approx 30\%$) are
slightly bigger than those of the randomly generated samples due to
the fact that our sample
contains only a few high column density systems.

To definitively answer the question of dust bias we would ideally hope
to at least approach the total redshift path searched by optical
surveys in order to make valid comparisons.  In the combined
radio-selected sample, the total cosmological redshift path searched
was $\Delta X\ = 317.11$.  Compared with the total redshift path
surveyed by the latest large optical survey, SDSS-DR3 with $\Delta X =
7333.2$, we are still more than an order of magnitude smaller.

The combined radio-selected central value of $\momg$ is slightly
higher than all of the optically-selected values, as plotted in
Figure~\ref{fig:omega}.  However, when considering the 1$\sigma$ lower
limits of the radio-selected values of $\momg$, no difference between
the magnitude-limited sample and the radio-selected samples can be
ascertained.  If we ignore the possibility of evolution in $\momg$ we
can compare the entire SDSS-DR3 optically-selected survey over the
complete redshift range with the radio-selected value.  Excluding the
P\'eroux point, the SDSS-DR3 value taken over one redshift bin, $z$ =
[2.2, 5.5], gives $\momg$ = 0.82$_{-0.05}^{+0.05}$ $\times 10^{-3}$.
Comparing this value with the 1$\sigma$ lower limit of the combined
radio-selected value of $\momg^{low} = 0.77 \times 10^{-3}$ over the
range $z$ = [1.51, 4.4], we see excellent agreement.  This agreement in
\omg\ between the radio and optically-selected surveys for \dlas\ is
the best evidence for our conclusion that dust bias does not have a
major effect on the results of optically-selected surveys.

\acknowledgments
A.M.W., R.A.J., \& J.X.P. are partially supported by
the National Science Foundation grant AST-0307824.

E.G. acknowledges support from the National Science Foundation under
grant AST-0201667, an NSF Astronomy
and Astrophysics Postdoctoral Fellowship (AAPF).

\end{document}